\documentclass[prl,10pt,preprint,twocolumn]{revtex4}
\usepackage{amsfonts}
\usepackage{amsmath}
\usepackage{amssymb}
\usepackage{graphicx}
\begin{document}
\preprint{Solid State Communications, Vol. 44, No. 3, pp. 387-389,
1982. 0038-1098/82/390387-03\$03.00/0} \preprint{Printed in Great
Britain Pergamon Press Ltd}

\title{THE EFFECTIVE SPIN HAMILTONIAN AND PHASE SEPARATION
INSTABILITY OF THE ALMOST HALF-FILLED HUBBARD MODEL AND
NARROW-BAND {\it s-f} MODEL}
\author{M.I. Auslender}
\author{M.I. Katsnelson}
\affiliation{{\em Institute for Metal Physics, Ural Scientific
Centre, Sverdlovsk, U.S.S.R.}}

\received{1 April 1982 by B. Muhlschlegel}

\begin{abstract}
\noindent The effective spin Hamiltonian is constructed in the
framework of the almost half-filled Hubbard model on the Cayley
tree by means of functional integral technique with the use of
static approximation. The system in the ground state appears to be
consisting of the ferromagnetic metallic domains and the
antiferromagnetic insulating one sprovided that the concentration
of excess electrons (or holes) does not exceed some critical
value. The connection between the Hubbard model and the {\it s-f} model
is stated.
\end{abstract}
\maketitle
\noindent TO DESCRIBE the narrow-band magnetic
semiconductdrs, one uses in general two models: (1) the Hubbard
model with the Hamiltonian
\begin{equation}
\mathcal{H}_{H}=\sum_{ij\sigma}\left(
t_{ij}-\mu\delta_{ij}\right)
c_{i\sigma}^{\dag}c_{j\sigma}+u\sum_{i}n_{i\uparrow}n_{i\downarrow
}\label{Hubbard}
\end{equation}
where $c_{i\sigma}^{\dag}$, $c_{i\sigma}$ are the creation and
annihilation electron operators on the site $i$ with the spin
projection $\sigma$, $n_{i\sigma}$ $=$
$c_{i\sigma}^{\dag}c_{i\sigma}$, $\mu$ is the chemical potential,
$u\gg|t_{ij}|$, an average number of electrons per site
$N_{e}/N=1+c$ ($c>0$) being close to unity; and (2) the {\it s-f}
model
\begin{align}
& \mathcal{H}_{s-f}=\sum_{ij\sigma}\left(
t_{ij}-\mu\delta_{ij}\right)
c_{i\sigma}^{\dag}c_{j\sigma}-I\sum_{i\sigma\sigma^{\prime}}\mathbf{S}
_{i}\cdot
c_{i\sigma}^{\dag}\boldsymbol{\sigma}_{\sigma\sigma^{\prime}
}c_{i\sigma^{\prime}}\nonumber \\ & + \mathcal{H}_{f}\label{s-f}
\end{align}
where $\mathbf{S}_{i}$ is the \textit{f}-spin operator, $\boldsymbol{\sigma}$
is the Pauli matrices, $\mathcal{H}_{f}$ is the Hamiltonian of direct
\textit{f}-spin interactions, $I$ is the exchange parameter, $\left\vert
I\right\vert \gg|t_{ij}|$, $N_{e}/N\ll1$. The first model describes the
situation when the conduction electrons belong to the same band as the
electrons forming the localized moments. On the other hand the Hamiltonian
(\ref{s-f}) is applied when these electrons are in the different bands
(\textit{s} and \textit{f}). Nevertheless the models (\ref{Hubbard}) and
(\ref{s-f}) are equivalent under certain approximations as it will be proved
further (Nagaev was the first who emphasized this feature \cite{nagaev}).
Therefore we restrict ourselves to consideration of the Hubbard model.

When both $c$ and $|t_{ij}|/u$ are small the competition of
tendencies to ferro- and antiferromagnetic orderings takes place
\cite{nagaoka,brink-rice,visscher} which leads to the phase
separation instability \cite{brink-rice,visscher}. So far this
effect has been treated qualitatively (see also
\cite{nagaev,mott}). We shall construct here the effective spin
Hamiltonian for the systems under consideration which will give
the possibility to describe this instability as the first-order
phase transition.

We proceed with the representation of the Hubbard model partition
function $Z$ in terms of the functional integral over the
fluctuating vector fields $\left\{
\boldsymbol{\varepsilon}_{i}\left(  \tau\right)  \right\}$
conjugated to the spin density (see reviews \cite{morya,morandi}):
\begin{equation}
Z = \int\prod\limits_{i}
\mathcal{D}\boldsymbol{\varepsilon}_{i}\left(  \tau\right)
\exp\left[-\frac{3}{2u}\int_{0}^{\beta}\boldsymbol{\varepsilon}_{i}^{2}\left(
\tau\right)  d\tau\right]  Q\left\{
\boldsymbol{\varepsilon}_{i}\left( \tau\right) \right\}
\label{Pf0}
\end{equation}
where $\beta^{-1}=T$ is the temperature ($\hbar=1$, $k=1$),
$Q\left\{ \boldsymbol{\varepsilon}_{i}\left(\tau\right) \right\} $
is the partition function of free electrons in the external
magnetic field $\boldsymbol{\varepsilon}_{i}\left( \tau\right)  $.
Then we use a common ``static approximation" \cite{morya,morandi},
i.e. we replace $\boldsymbol{\varepsilon}_{i}\left( \tau\right) $
by its mean value over $\left( 0,\beta\right)  $ -
$\boldsymbol{\varepsilon}_{i}$. At $t_{ij}$ $=0$, $Q\left\{
\boldsymbol{\varepsilon}_{i}\left(  \tau\right)  \right\}  $ can
be calculated exactly. Considering formally the band energy term
in equation (\ref{Hubbard}) as a perturbation we get
\begin{align}
& Z = \left(\frac{6\beta}{\pi u}\right)^{3N/2}\exp\left(N\beta
\widetilde{\mu}\right)\int\prod\limits_{i}
d\boldsymbol{\varepsilon}_{i}\exp\left[-\left(3\beta
\boldsymbol{\varepsilon}_{i}^{2}/2u\right)\right] \nonumber
\\& \times\left[\cosh\left(\beta\widetilde{\mu}\right)
+ \cosh\left( \beta\varepsilon_{i}\right)
\right]\exp\left(-\beta\Phi\left\{
\boldsymbol{\varepsilon}_{j}\right\}  \right) ,\nonumber
\\ & \Phi\left\{  \boldsymbol{\varepsilon}_{j}\right\}
=-T\mbox{Tr}\ln\left(I-tG\right),\label{Pf1}
\end{align}
where $\widetilde{\mu}= \mu-u/2$, $t$, $G$ are the operators with
matrix elements:
\begin{align}
t_{ij}^{\sigma\sigma^{\prime}}\left(  \tau,\tau^{\prime}\right)    &
=t_{ij}\delta_{\sigma\sigma^{\prime}}\delta\left(  \tau-\tau^{\prime}\right)
;\nonumber\\
G_{ij}^{\sigma\sigma^{\prime}}\left(  \tau,\tau^{\prime}\right) &
=\delta_{ij}T\sum_{n}G_{j}^{\sigma\sigma^{\prime}}\left(
i\omega_{n}\right) e^{-i\omega_{n}\left( \tau-\tau^{\prime}\right)
}\nonumber
\end{align}
respectively, $\omega_{n}=\pi T\left(  2n+1\right)  $, $n=0,\pm1,\pm2,\ldots$,%
\begin{align}
G_{j}^{\sigma\sigma^{\prime}}\left(  i\omega_{n}\right)    & =\left(
i\omega_{n}+\widetilde{\mu}-\boldsymbol{\varepsilon}_{j}\cdot
\boldsymbol{\sigma}\right)  _{\sigma\sigma^{\prime}}^{-1}\nonumber\\
& =\frac{i\omega_{n}+\widetilde{\mu}+\boldsymbol{\varepsilon}_{j}%
\cdot\boldsymbol{\sigma}_{\sigma\sigma^{\prime}}}{\left(  i\omega
_{n}+\widetilde{\mu}\right)
^{2}-\varepsilon_{j}^{2}}\label{locGF-zo}
\end{align}
is the Green function of zeroth order approximation in $t_{ij}$. $\Phi
\equiv\Phi\left\{  \boldsymbol{\varepsilon}_{j}\right\}  $ is the functional
of the spin interaction energy. When calculating the integral over the modulus
of $\boldsymbol{\varepsilon}_{j}$, we can apply the saddle point method since
$u\gg T$, $\left\vert t_{ij}\right\vert $: $\boldsymbol{\varepsilon}_{j}%
\simeq\epsilon_{0}\mathbf{n}_{j}$ ($\epsilon_{0}=u/3$, $\left\vert
\mathbf{n}_{j}\right\vert =1$).  Then $\Phi$ appears to be the functional only
of unit vectors $\mathbf{n}_{j}$. The formal perturbation expansion of
$\ \Phi\left\{  \mathbf{n}_{j}\right\}  $ with respect to $t_{ij}$ may be
obtained in the usual way. It turns out that the true parameter of the
expansion is $\left\vert t_{ij}\right\vert /\left\vert \epsilon_{0}%
-\widetilde{\mu}\right\vert $. It is of order of $\left\vert
t_{ij}\right\vert /\epsilon_{0}\ll1$ at $c=0$. But if $c$ is small
but finite, $c\gg\exp(-\beta\epsilon_{0})$, we have
$\widetilde{\mu}$ $=\epsilon_{0}+T\ln c/(1-c)$ in the zeroth order
approximation with respect to $t_{ij}$, and therefore the
perturbation expansion fails. In the case of
$c\neq0$ we use the representation of functional $\Phi\left\{  \mathbf{n}%
_{j}\right\}  $ through the exact Green function $\mathcal{G}_{ij}\left(
\omega|\lambda\right)  $:%
\begin{align}
& \Phi  =-\frac{1}{\pi}\int_{-\infty}^{+\infty}d\omega f\left(
\omega -\widetilde{\mu}\right)
\sum_{i}\operatorname{Im}\Lambda_{i}\left(
\omega^{+}\right),\;\omega^{+}=\omega+i0,\nonumber
\\& \Lambda_{i}\left(  \omega\right) =\mbox{Tr}_{\sigma}\int_{0}^{1}
d\lambda\sum_{j}t_{ij}\mathcal{G}_{ij}\left(  \omega|\lambda\right)
= \nonumber\\
& \int_{0}^{1}\frac{d\lambda}{\lambda}\mbox{Tr}_{\sigma}\left[
\left(
\omega-\epsilon_{0}\mathbf{n}_{i}\cdot\boldsymbol{\sigma}\right)
\mathcal{G}_{ii}\left(  \omega|\lambda\right)  -1\right]  ,\label{Fi}%
\end{align}
in which $f(x)=(\exp\beta x+1)^{-1}$ and $\mathcal{G}_{ij}\left(
\omega|\lambda\right)  $ satisfies the following equation%
\begin{equation}
\left(
\omega-\epsilon_{0}\mathbf{n}_{i}\cdot\boldsymbol{\sigma}\right)
\mathcal{G}_{ij}\left(\omega|\lambda\right)
-\sum_{j^{\prime}}\lambda
t_{ij^{\prime}}\mathcal{G}_{j^{\prime}j}\left(
\omega|\lambda\right) =\delta_{ij}.\label{GF-eq}
\end{equation}
For the \textit{s}-\textit{f }model with classical
\textit{f}-spins $\mathbf{S}_{j}=S\mathbf{n}_{j}$ there is the
correspondence with equations (6) and (7) if one replaces
$\epsilon_{0}$ $\rightarrow\left\vert I\right\vert S$. This proves
rigorously the equivalence of the narrow-band Hubbard model in the
static approximation to the \textit{s}-\textit{f } model with
classical \textit{f}-spins and $\mathcal{H}_{f}=0$.

Because the equation (\ref{Fi}) contains the diagonal Green functions only we
try to derive a closed equation for it. It is convenient to introduce the
locator self-energy $\mathcal{L}_{i}\left(  \omega|\lambda\right)  $ by the
relation%
\begin{equation}
\mathcal{G}_{ii}^{-1}\left(  \omega|\lambda\right)  =\omega-\epsilon
_{0}\mathbf{n}_{i}\cdot\boldsymbol{\sigma}-\mathcal{L}_{i}\left(
\omega|\lambda\right)  \label{L}%
\end{equation}
The following expression for $\mathcal{L}_{i}\left(
\omega|\lambda\right)  $ was proposed in \cite{abou-chacra} (for
the case of electrons moving in disordered media)
\begin{align}
& \mathcal{L}_{i}\left(  \omega|\lambda\right)
=\lambda^{2}\sum_{j}\left\vert t_{ij}\right\vert
^{2}\mathcal{G}_{jj}\left(  \omega|\lambda\right) = \nonumber
\\& \left(
\lambda t\right)  ^{2}\sum_{\boldsymbol{\delta}}\mathcal{G}%
_{i+\boldsymbol{\delta,}i\boldsymbol{+\delta}}\left(
\omega|\lambda\right) \label{LvsG}
\end{align}
where $\boldsymbol{\delta}$ labels the nearest neighbours (the
nearest neighbour approximation is assumed for $t_{ij}$). Equation
(\ref{LvsG}) can be obtained from the second-order term of the
perturbation theory for $\mathcal{L} _{i}\left(
\omega|\lambda\right) $ by replacing $G_{j}\left(  \omega
|\lambda\right) \rightarrow\mathcal{G}_{jj}\left(
\omega|\lambda\right) $. It turns out to be exact on the Cayley
tree \cite{abou-chacra}. Both in the ferromagnetic (FM) and the
antiferromagnetic (AFM) cases $\mathbf{n}_{i}
\cdot\mathbf{n}_{i+\boldsymbol{\delta}}$ does not depend on $i$.
Therefore we assume that the function $\cos\theta_{i}$
$=(1/z)\sum_{\boldsymbol{\delta}
}\mathbf{n}_{i}\cdot\mathbf{n}_{i+\boldsymbol{\delta}}$ ($z$ is
the coordination number) is a smooth function of site $i$. In this
approximation and on account that $\left\vert t\right\vert
/\epsilon_{0}\ll1$ the matrix equation (\ref{LvsG}) can be solved.
The solution to equation (\ref{LvsG})
gives%
\begin{align}
\Lambda_{i}\left(  \omega\right)    & =2\int_{0}^{1}\frac{d\lambda}{\lambda
}\left[  \left(  \omega^{2}-\epsilon_{0}^{2}\right)  X_{i}\left(
\omega|\lambda\right)  -1\right]  \nonumber\\
X_{i}\left(  \omega|\lambda\right)    & =2\left\{  \omega^{2}-\epsilon_{0}%
^{2}+\left[  \left(  \omega^{2}-\epsilon_{0}^{2}\right)  ^{2}-8z\left(
\lambda t\right)  ^{2}\right.  \right.  \nonumber\\
& \left.  \left.  \times\left(  \omega^{2}+\epsilon_{0}^{2}\cos\theta
_{i}\right)  \right]  ^{1/2}\right\}  ^{-1}\label{L-fin}%
\end{align}
(the branch of square root is chosen by the condition $X_{i}\left(
\omega|\lambda\right)  \rightarrow\left(  \omega^{2}-\epsilon_{0}^{2}\right)
^{-1}$ at $t\rightarrow0$). Substituting equation (\ref{L-fin}) to equation
(\ref{Fi}), taking into account the smallness of $c$, $\left\vert t\right\vert
/\epsilon_{0}$ and integrating over $\lambda$ and $\omega$ we obtain%
\begin{align}
& \Phi\left\{  \mathbf{n}_{j}\right\}
\simeq\frac{zt^{2}}{2\epsilon_{0}
}\sum_{i}\cos\theta_{i}-\frac{8\sqrt{z}\left\vert t\right\vert
}{15\pi} \sum_{i}\cos\frac{\theta_{i}}{2}\Theta\left(
\widetilde{\mu}-E_{i}\right) \nonumber
\\& \times\left(  \frac{\widetilde{\mu}-E_{i}}{\sqrt{z}\left\vert t\right\vert
\cos\frac{\theta_{i}}{2}}\right)  ^{5/2}+\mbox{const};  \nonumber
\\ & \Theta\left( x\right) =\left\{
\begin{array}
[c]{c} 1,\,x>0,\\ 0,\,x<0,
\end{array}
\right.  \label{Fi-fin}
\end{align}
where
\begin{equation}
E_{i}=\epsilon_{0}-2\sqrt{z}\left\vert t\right\vert \cos\frac{\theta_{i}}%
{2}+\frac{zt^{2}}{\epsilon_{0}}\left(  1-\cos\theta_{i}\right)  .\label{E}%
\end{equation}
$\Phi\left\{  \mathbf{n}_{j}\right\}$ is desired free energy
functional. It is necessary to write down the equation for the
shifted chemical potential $\widetilde{\mu}$ in addition to
equations (\ref{Fi-fin}) and (\ref{E}). It reads
\begin{align}
& c  =\left\langle c\left\{  \mathbf{n}_{j}\right\} \right\rangle
_{\Phi },\;c\left\{  \mathbf{n}_{j}\right\} =\frac{4}{3\pi
N}\sum_{i}\left( \frac{\widetilde{\mu}-E_{i}}{\sqrt{z}\left\vert
t\right\vert \cos\frac
{\theta_{i}}{2}}\right)  ^{3/2}\nonumber\\
& \times\Theta\left(  \widetilde{\mu}-E_{i}\right)  ;\nonumber
\\ & \left\langle A\left\{  \mathbf{n}_{j}\right\}\right\rangle
_{\Phi}=\frac{ \int \prod\limits_{i} d\mathbf{n}_{i}A\left\{
\mathbf{n}_{j}\right\}  \exp\left(  -\beta \Phi\left\{
\mathbf{n}_{j}\right\}  \right)  }{\int \prod\limits_{i}
d\mathbf{n}_{j}\exp\left(  -\beta\Phi\left\{
\mathbf{n}_{j}\right\}  \right) }\label{mu-eq}
\end{align}
At $T=0$ the total energy of the system $E=-T\ln Z+\widetilde{\mu}N_{e}$
equals $E=\min(\Phi\left\{  \mathbf{n}_{j}\right\}  +Nc\widetilde{\mu}\left\{
\mathbf{n}_{j}\right\}  )$, $\widetilde{\mu}\left\{  \mathbf{n}_{j}\right\}  $
being the chemical potential for the fixed spin configuration when
$c\rightarrow c\left\{  \mathbf{n}_{j}\right\}  $.

We minimize the energy putting $\cos\frac{\theta_{i}}{2}$ $=1$ in the domain
of relative volume $x$ (FM phase) and $\cos\frac{\theta_{i}}{2}$ $=0$ in the
domain of relative volume $1-x$ (AFM phase). The optimal value of $x$ is
$x_{0}=c/c_{0}$,%
\begin{equation}
c_{0}=\left[  \frac{5}{2}\left(  \frac{4}{3\pi}\right)  ^{2/3}\frac{\sqrt
{z}\left\vert t\right\vert }{\epsilon_{0}}\right]  ^{3/5}.\label{c_0}%
\end{equation}
At $c<c_{0}$ the two-phase state obtained has the energy which is lower than
that of any homogeneous state. Apart from numerical factors, equation
(\ref{c_0}) agrees with the result of the qualitative consideration
\cite{visscher}. It can be seen from equation (\ref{mu-eq}) that all excess
electrons are in FM domains. At $c>c_{0}$ the homogeneous FM ordering is
favourable. Our Hamiltonian reduces to double-exchange Hamiltonian derived
earlier in \textit{s}-\textit{f} exchange model (see, e.g. \cite{anderson})
provided that $c_{0}\ll c\ll1$. It is worthwhile to mention that the Cayley
tree approximation gives the right value for the AFM indirect exchange
parameter but overestimates the energy of FM ordering in $\frac{1}{2}\sqrt{z}$ times.

The contribution of space derivatives of the function
$\cos\theta_i$ to the free energy functional has been calculated
also. It appears that the thickness of the boundary between FM and
AFM phases is of order of lattice constant. This matter will be
published elsewhere.

\end{document}